\documentclass{ajour}
\begin{document}
\thispagestyle{empty}

\newpage

%



\authorrunninghead{Qian}
\titlerunninghead{Mathematical Theory of Linear Irreversibility}



\def\bfE{\mbox{\boldmath$E$}}
\def\bfG{\mbox{\boldmath$G$}}
\def\bfR{\mbox{\boldmath$R$}}
\def\vecJ{\mbox{\boldmath$J$}}
\def\vecb{\mbox{\boldmath$b$}}
\def\vecPi{\mbox{\boldmath$\Pi$}}


\title{MATHEMATICAL FORMALISM FOR ISOTHERMAL 
	LINEAR IRREVERSIBILITY}


\author{Hong Qian}

\affil{Department of Applied Mathematics, Universityi of Washington, \\ 
Seattle, WA 98195-2420}

\email{qian@amath.washington.edu}


\abstract{
\large
\baselineskip = 0.27 in
We prove the equivalence among symmetricity, time reversibility,
and zero entropy production of the stationary solutions of linear
stochastic differential equations.  A sufficient and necessary
reversibility condition expressed in terms of the coefficients
of the equations is given.  The existence of a linear stationary
irreversible process is established.  Concerning reversibility,
we show that there is a contradistinction between any
1-dimensional stationary Gaussian process and stationary
Gaussian process of dimension $n>1$.  A concrete criterion for
differentiating stationarity and sweeping behavior is also
obtained.  The mathematical result is a natural generalization
of Einstein's fluctuation-dissipation relation, and provides a
rigorous basis for the isothermal irreversibility in a linear 
regime which is the basis for applying Onsager's theory to 
macromolecules in aqueous solution.}


\begin{article}
\section{Introduction}

	The stochastic differential equation 
\begin{equation}
        \frac{dx}{dt} = \vecb(x) + \Gamma\xi(t),  \hspace{0.5cm}
			x\in\bfR^n
\label{NLSDE}
\end{equation}
with $\Gamma$ being a nonsingular matrix and $\xi(t)$ 
being the ``derivative'' of a $n$-dimensional Wiener process, 
has wide applications in science and engineering \cite{Ok}. 
One standard method for attacking this equation is by finding
the fundamental solution to its corresponding Fokker-Planck 
(Kolmogorov forward) equation
\begin{equation}
    \frac{\partial P}{\partial t} = 
		\nabla\cdot\left(\frac{1}{2}A\nabla P 
					- \vecb(x)P\right), 
		\hspace{0.5cm}
		(A = \Gamma\Gamma^T)
\label{NLFPE}
\end{equation}
a parabolic equation for the transition probability $P(x,t|x^0)$
defined on the entire $\bfR^n$ with the integrability condition
\begin{equation}
        \int_{\bfR^n} |P(x,t)| dx < +\infty.
\label{L1}
\end{equation}
That is $P\in L^1[\bfR ^n]$.  With this condition, it has
been difficult to obtain detailed, rigorous understanding 
of the partial differential equation (PDE) in (\ref{NLFPE}).
However, for a class of PDE satisfying the regular conditions 
for the Cauchy problem, which include linear $\vecb(x)$, 
the existence and uniquess of a positive fundamental solution
is guaranteed.  Furthermore, the existence and uniqueness of
the solution to the Cauchy problem with initial data
$f(x)$ ($|f(x)| \le ce^{\alpha |x|^2}$ with constant $c>0$
and $\alpha >0$) is proven.  The above mentioned regular 
condition for Cauchy problem and its consequences can be 
found in \cite{LM}.  In addition, a result on the 
nonexistence of a stationary solution in $L^1$, 
called sweeping, is also obtained.  This is stated as follows 
\cite{LM}: 

\vskip 0.3cm 
\begin{quote}{\em
Assume that the coefficients of equation (\ref{NLFPE}) are
regular for the Cauchy problem.  Further assume that all 
stationary nonnegative solution of (\ref{NLFPE}) and 
(\ref{L1}) are of the
form $cu(x)$ where $u(x)>0$ almost everywhere and 
$c$ is a nonnegative constant.  Then the solution of
(\ref{NLFPE}) is either asymptotically stable or sweeping.
Asymptotic stability occurs when 
\[             I \equiv \int_{\bfR^n} u(x) dx < \infty  \]
and sweeping when $I=\infty.$}
\end{quote} 

\vskip 0.3cm \noindent
The condition in this sweeping theorem is satisfied 
because of the uniqueness and positivity of the fundamental 
solution to (\ref{NLFPE}) and the corresponding Cauchy
problem \cite{LM}.  No criterion has been given to assert when 
the sweeping or asymptotically stable behavior indeed occurr.  

	In this paper, we show that this problem can be 
completely solved for linear $\vecb(x)$.  A sufficient and 
necessary condition for the sweeping behavior is obtained. 
Within the nonsweeping case, the stationary processes 
can be further classified as either reversible or irreversible.
A series of sufficient and necessary conditions, 
and a criterion based on the coeffients of 
(\ref{NLSDE}) are derived for this classification. 
The distributions concerned are all Gaussian.  In
the case of $n=1$, Gaussian stationary processes are 
necessarily reversible \cite{Weiss}. Part of our results,
thus, may be regarded as a further development of this theorem 
for arbitrary $n>1$. 

	The stochastic differential equation given in (\ref{NLSDE}) 
defines a Markovian process known as diffusion.  This 
theory has been a well accepted mathematical model for the 
statistical behavior of molecular systems, following the work 
of Einstein, Ornstein and Uhlenbeck, Onsager, Keizer, and 
other physcists \cite{Wax,RRF,Ke,Qian2,Qian3}.  
One of the central pieces in the physicists' work is the 
fluctuation-dissipation relation. Until now, however, the 
role of this deep relationship has not led to any mathematical 
conclusion in the theory of stochastic differential equation. 

	It turns out, the fluctuation-dissipation relation 
naturally emerges from the mathematical results of the 
present paper. We will show that the standard 
fluctuation-dissipation relation is a necessary but not 
a sufficient condition for time-reversibility. Our reversibility 
condition, on the other hand, can be viewed as a 
stronger form of fluctuation-dissipation relation which also 
has a closer resemblance to Einstein's original
one.  Time-reversibility is an important concept in which the 
physicists are interested.  In connection to 
this end, we introduce the concept of
entropy production rate, derive its analytical
expression and prove that in a stationary process
it is zero if and only if the process is reversible.
Finally, we provide the condition under which a 
stationary irreversible Gaussian process exists.

\section{Thermodynamics, Heat Dissipation, and 
Entropy Production Rates} 

	In recent years, with the increasing number of applications
of the stochastic model (\ref{NLSDE}) or (\ref{NLFPE}) to a wide 
range of macromolecular processes \cite{Hill,JAP,Qian4,Qian5}, it 
has become evident that an axiomatic isothermal thermodynamic 
formalism can be established based on the stochastic differential 
equation.  In this section, we give a brief introduction of this 
emerging theoretical framework we called {\it stochastic 
macromolecular mechanics} \cite{Qian5}.

	Following Lebowitz and Spohn \cite{LS}, we introduce the 
instananeous heat dissipation functional $W_t$ \cite{Qian6}:
\begin{equation}
        dW_t = (A/2)^{-1}\vecb(X_t)\circ dX_t 
              = 2A^{-1}\vecb(X_t)\cdot dX_t 
	      +\nabla\cdot\vecb(X_t) dt
\label{W}
\end{equation}
where $\circ$ denotes the integral in the Stratonovich sense.
According to this definition, the heat dissipation, the 
left-hand-side of (\ref{W}), is equal to the work done by 
the system, the right-hand-side of (\ref{W}). 
The work is the product of force and displacement; 
the force is the product of the frictional coefficient 
$(2A^{-1})$ and velocity $\vecb(x)$.  This is the law of energy 
conservation.  It then follows that the mean rate of the heat 
dissipation 
\begin{equation}
 \textrm{hdr} = \frac{d}{dt}E\left[W_t\right] 
              = \int_{\bfR^n} 2(A^{-1}\vecb(x))\cdot \vecJ dx
\end{equation}
in which 
\[    \vecJ = -\frac{1}{2}A\nabla P(x,t)+\vecb(x)P(x,t)  \]
denoting the probability flux in (\ref{NLFPE}): 
$\partial P(x,t)/\partial t$ = $-\nabla\cdot\vecJ$, and the 
probability density $P(x,t)$ is the solution to (\ref{NLFPE}).
In mechanical terms, the mean heat dissipation rate is the 
product of force $(2A^{-1}\vecb(x))$ and flux ($\vecJ$) \cite{Ke}.

	The thermodynamic force of Onsager is different from
the mechanical force $2A^{-1}\vecb(x)$.  In terms of 
(\ref{NLFPE}), the Onsager's thermodynamic force $\vecPi(x)$,
called affinity by chemists, is \cite{Hill}
\begin{eqnarray}
     \vecPi(x)\cdot dx &=& \log\left(\frac{P(x+dx,t|x)P(x)}
			{P(x,t|x+dx)P(x+dx)}\right)
\\[5pt]
	&\approx& \left\{2A^{-1}\vecb(x)-\nabla\log P(x)\right\}
							\cdot dx
\nonumber
\end{eqnarray}
in which $\approx$ is valid for small $dx$.  Therefore, 
the thermodynamic force 
$\vecPi$ and flux $\vecJ$ are related by $\vecJ = \frac{P}{2}A\vecPi$,
in which the matrix $A$ is symmetric.  This relation reflects the 
Onsager's reciprocal relations \cite{On}.  When applying (\ref{NLSDE}) 
to molecular motors, a subspace of $x$ represents the internal motion 
of a macromolecule, and the remains of the $x$ represents the external 
movement of the entire macromolecule \cite{Qian}.  Hence the relation
between $\vecJ$ and $\vecPi$ reflects a certain symmetry between the 
chemical flux, the mechanical movement, the chemical energy source, 
and the mechanical work in the energy convertion. 

	We now introduce another two important thermodynamic 
quantities: the entropy and entropy production rate (epr).  The 
resulting analytical expression will then be used as a rigorous 
definition for the remains of the paper.  

	We use the well-known definition for entropy
\begin{equation}
           e[P] = -\int_{\bfR^n} P(x,t)\log P(x,t) dx
\end{equation}
which is a functional of the probability density $P(x,t)$, the 
solution of (\ref{NLFPE}).  

The rate of the increase of 
entropy is
\begin{eqnarray}
  \dot{e}[P] &=& \int_{\bfR^n} (\log P + 1)\nabla\cdot\vecJ\
         dx    \nonumber \\
    &=& -\int_{\bfR^n} P^{-1}(\nabla P-2A^{-1}\vecb(x)P)\cdot\vecJ\ dx
        -\int_{\bfR^n} 2A^{-1}\vecb(x)\cdot\vecJ\ dx 
		\nonumber \\
    &=& \int_{\bfR^n} \vecPi\cdot\vecJ dx 
			- \int_{\bfR^n} 2A^{-1}\vecb(x)\cdot\vecJ dx 
			\label{depr} \\
    &=& \textrm{epr} - \textrm{hdr}.    \nonumber
\end{eqnarray}
In the derivation, we used equation (\ref{NLFPE}) and 
integration by parts, assumed no flux 
boundary condition and matrix $A$ being nonsingular.  It is
meaningful from the thermodynamic point of view to identify the
first term in (\ref{depr}) with the entropy production rate which
equals to the product of thermodynamic force and flux, and the 
second term is exactly the heat dissipation rate.  In a time 
independent stationary state, the $\dot{e} = 0$, and the entropy 
production is balanced by the heat dissipation.  This is the case
for an isothermal nonequilibrium steady-state.

	For a single macromolecule immersed in a fluid with 
constant temperature, our $epr$ in (\ref{depr}) is related to 
Onsager's dissipation function $epr/2=\Phi$, our $hdr$ is his 
$\dot{S}^*$, and our $\dot{e}$ is his $\dot{S}$.  Therefore, 
(\ref{depr}) corresponds to $\dot{S}=2\Phi-\dot{S}^*$ in which 
$2\Phi$ is a ``source'' term, entropy production, $\dot{S}^*$ 
is the heat giving out by the system to the surrounding fluid, 
and $\dot{S}$ is the entropy change of the system proper.

	We now define the entropy production rate and time 
reversibility.
\begin{definition}
The entropy production rate, $epr$, of a stationary diffusion
process defined by (\ref{NLSDE}) is
\begin{equation}
     \frac{1}{2} \int (\nabla\log P(x)-2A^{-1}\vecb(x))^T
		A(\nabla\log P(x)-2A^{-1}\vecb(x))P(x)dx. 
\label{epr}
\end{equation}
\end{definition}
It is seen that the $epr$ is always non-negative.  This is the
second law of thermodynamics.  This paper is to establish a
mathematical relationship between $epr=0$ and the time
reversibility defined as follows:

\begin{definition}
A stationary stochastic process $\{x(t);t\in\bfR\}$ is time 
reversible if for every positive integer $m$ and every 
$t_1$,$t_2$,...,$t_m$ $\in \bfR$, the joint probability 
distribution
\[     P(x(t_1),x(t_2),...,x(t_m)) =
		 P(x(-t_1),x(-t_2),...,x(-t_m)).    \]
\end{definition}

	Eq. \ref{epr} also indicates that the $epr$
equals zero if and only if $2A^{-1}\vecb$ = $\nabla\log P$.  That 
is the force $2A^{-1}\vecb$ being conservative with a potential:
$2A^{-1}\vecb$ = -$\nabla U$.  For systems satisfying the potential
condition, the thermodynamic force $\vecPi$ also has a potential,
$\vecPi$ = -$\nabla(U(x)+\log P(x))$, whose expectation is precisely
the Helmholtz free energy: 
\begin{eqnarray}
          \Psi[P] &=&  E\left[U(x)+\log P(x)\right]
\nonumber\\
	  &=& \int_{\bfR^n} P(x,t)U(x)dx-e[P]. 
\end{eqnarray}
in which the first term is the mean internal energy.
It is easy to show that $\dot{\Psi}$ = $-epr \le 0$: the free energy 
increases and reaches its maximum at an equilibrium. It turns out 
that $\Psi$ is related to the relative entropy \cite{Qian7}, and 
it is well-known that relative entropy is a Lyapunov function
for Eq. \ref{NLFPE} \cite{LM}.  Finally, with the potential
condition, the heat dissipation functional $dW_t$ = $dU(X_t)$,
the internal energy fluctuation.  Hence, $W_t$ is stationary and 
its expectation and variance are the internal energy and heat 
capacity of a single macromolecule at thermal equilibrium.
A thermal equilibrium is necessarily time reversible.

\section{Linear Theory of Reversible Stationary Processes}

Let's consider the linear stochastic differential equation of order
$n$:
\begin{equation}
 \frac{d x}{dt} = -Bx + \Gamma\xi(t)
\label{ndlsde}
\end{equation}
where
\begin{equation}
x(t)=\left[ \begin{array}{c}
x_1(t)  \\
\vdots  \\
x_n(t)  \\ \end{array}\right],
\hspace{0.2cm}
B=\left[ \begin{array}{ccc}
b_{11}  &\cdots &b_{1n}\\
\vdots  &       &\vdots\\
b_{n1}  &\cdots &b_{nn}\\ \end{array}\right],
\hspace{0.2cm}
\Gamma=\left[ \begin{array}{ccc}
\gamma_{11}     &\cdots &\gamma_{1n}\\
\vdots  & 	&\vdots\\
\gamma_{n1}  &\cdots	&\gamma_{nn}\\ \end{array}\right]
\end{equation} 
is nonsingular, and
\begin{equation} 
\xi(t)=\left[ \begin{array}{c} 
\xi_1(t)	\\
\vdots  	\\
\xi_n(t)  	\\ \end{array}\right] 
\end{equation}
is the ``white noise'' vector which should be considered, from
a mathematical point of view, as the derivative of a $n$-dimensional
Wiener process. Formally, we have $E[\xi(t)\xi^T(t')]$ =
$\delta(t-t')I$ where $I$ is the identity matrix.

	Heuristically speaking, near any stable fixed 
point of $\vecb(x)=0$ in a nonlinear (\ref{NLSDE}) there is a
linear approximation in the form of (\ref{ndlsde}). Therefore, 
the mathematical analysis presented in the present paper can 
be considered as the linear theory near such fixed point for the
nonlinear system.  The linear approximation leads to a Gaussian
process defined by (\ref{ndlsde}). 

	The probability density of the solution of (\ref{ndlsde}),
$P(x,t)$, satisfies the Fokker-Planck equation
\begin{equation}
    \frac{\partial P}{\partial t} = 
		\nabla\cdot\left(\frac{1}{2}A\nabla P 
			+ BxP\right) 
\label{FPE}
\end{equation}
where $A=\Gamma\Gamma^T$, which is positive definite since
we assume $\Gamma$ is nonsingular.
Hence (\ref{FPE}) is uniformly parabolic.

\begin{theorem}
\label{basic}
The following five statements about (\ref{ndlsde}) 
are equivalent:

(i) Its $B$ and $\Gamma\Gamma^T=A$ satisfy a symmetry condition 
$A^{-1}B=(A^{-1}B)^T$ which is positive definite;

(ii) It has a stationary Gaussian solution with a
symmetric two-time covariance matrix $R(t,t')$
=$E[x(t)x^T(t')]$;

(iii) It defines a time reversible stationary process;

(iv) Its corresponding elliptic operator (the right-hand-side of 
(\ref{FPE})) is symmetric with respect to a positive function
$w^{-1}(x)$, $w(x) \in L^1$; 

(v) Its stationary process has zero entropy production rate $(epr)$.
\end{theorem}
\vskip 0.3cm 

	{\bf Remark 1:} According to \cite{LM}, the stationary process
mentioned above is unique.

	{\bf Remark 2:} The statement $(i)$ is also equivalent to 
$A^{-1}Bx$ having a potential function $U(x)$ = 
$\frac{1}{2}x^TA^{-1}Bx$ and 
$\int_{\bfR^n} e^{-U(x)}dx$ $< \infty$.  The reversibility 
under this potential condition has been announced 
for the general nonlinear equation (\ref{NLSDE}) but without
proof \cite{QQ,Qian}, and discussed in \cite{GQW,QW} for 
the equation on compact manifold.  In this paper, we 
provide a simple proof for the linear case.

\vskip 0.3cm
\begin{proof}
{\bf (i) $\Rightarrow$ (ii).} 
The symmetry and positive definiteness of $A^{-1}B$=
$\Gamma^{-T}\Gamma^{-1}B$ implies that $\Gamma^{-1}B\Gamma$
is also symmetric and positive definite. Hence there exists an 
orthogonal matrix $Q$, $Q^T=Q^{-1}$, such that 
$\Gamma^{-1}B\Gamma$ = $Q\Lambda Q^{-1}$ where $\Lambda$ 
is a diagonal matrix with all positive eigenvalues
$\lambda_1, \lambda_2,...,\lambda_n$.  Thus
\begin{equation}
      B = \Gamma Q \Lambda Q^{-1}\Gamma^{-1}. 
\label{BBB}
\end{equation}
which has the same $\lambda$'s as eigenvalues.  Hence it is
nonsingular.  Substituting (\ref{BBB}) into (\ref{ndlsde}), 
we have
\begin{equation}
  \frac{dy}{dt} = -\Lambda y + \zeta(t)
\label{sep}
\end{equation}
where $y=Q^T\Gamma^{-1}x$, $\zeta(t)=Q^T\xi(t)$, and 
$E[\zeta(t)\zeta^T(t')]$ = $E[\xi(t)\xi^T(t')]$ = $\delta(t-t')I$. 
(\ref{sep}) is completely diagonalized.  Its solution are 
$n$ Ornstein-Uhlenbeck processes each with transition probability
\begin{equation}
  P(y_i,t|y_i^0) =\frac{1}{\sqrt{2\pi}\omega_i(t)}\ 
\exp\left[-\frac{(y_i-y_i^0e^{-\lambda_it})^2}{2\omega_i^2(t)}\right]
\end{equation} 
and stationary distribution
\begin{equation}
   P(y_i) = \frac{1}{\sqrt{2\pi}\omega_i(\infty)}\ 
\exp\left[-\frac{y_i^2}{2\omega_i^2(\infty)}\right]
\end{equation}
where 
\[     \omega_i^2(t) =\frac{1}{2\lambda_i}
			\left(1-e^{-2\lambda_i t}\right).      \]
Furthermore, the two-time covariance matrix for the stationary
process is
\begin{equation}
    E[y(t)y^T(t')] = \frac{1}{2}e^{-\Lambda |t-t'|}
				\Lambda^{-1}. 
\label{2time}
\end{equation}
Therefore, by substituting $y=Q^T\Gamma^{-1}x$ back into the
above equations, we have
\begin{equation}
 P(x,t|x^0) =\frac{\|E\|}{\sqrt{2\pi}}\ 
 \exp\left[-\frac{1}{2}(x-e^{-Bt}x^0)^TE(t)(x-e^{-Bt}x^0)\right]
\label{tb}
\end{equation}
in which matrix
\[   E^{-1}(t) 
   = \frac{1}{2}\left(I-e^{-2Bt}\right)B^{-1}\Gamma\Gamma^T.   \]
The stationary covariance $\Xi$ = $E^{-1}(\infty)$=$\frac{1}{2}B^{-1}A$
gives a stationary Gaussian distribution.  Combining with the
transition probability in (\ref{tb}) and Markovian properties, a 
Gaussian stationary process with covariance $\Xi$ is obtained
for (\ref{ndlsde}).  

	Finally, substituting $y=Q^T\Gamma^{-1}x$ back into
(\ref{2time}), the stationary two-time covariance matrix
\[   R(t,t') = E[x(t)x^T(t')] 
			= \frac{1}{2} e^{-B|t-t'|}B^{-1}A.     \]
We now show that $R(t,t')$ is symmetric.  
$A^{-1}B=B^TA^{-1}$ immediately leads to $BA=AB^T$, which 
gives:
\[     B^nA = B^{n-1}AB^T = B^{n-2}A(B^2)^T=...=A(B^n)^T.  \]
That is $B^nA = (B^nA)^T$ for any interger $n$.  Therefore
matrix $R(t,t')$ is symmetric.

\vskip 0.3cm
{\bf (ii) $\Rightarrow$ (iii).} 
	A Gaussian process $x(t)$ is completely determined by its
expectation $E[x(t)]$ and covariance $E[x(t)x^T(t')]$.  A
stationary Markovian Gaussian process is completely determined by 
the joint distribution $P(x_1,x_0)$.  The Gaussian
$P(x_1,x_0)$ is determined by all its first and second moments, 
among which are the two-time covariance matrix $R(t,0)$ =
$E[x(t)x_0^T]$. Therefore, the symmetric matrix $R(t,t')$=
$R(|t-t'|,0)$ leads to $P(x_1,x_0)=P(x_0,x_1)$;
and with the Markovian property the stationary Gaussian 
process is then time reversible.\footnote{It is interesting
to compare the cases of $n=1$ and $n>1$: It is known in 
statistics literature \cite{Weiss} that 
any 1-dimensional stationary Gaussian process is necessarily
time reversible, and any n-dimensional stationary Gaussian
process can be realized as a solution of (\ref{ndlsde}) \cite{VanK}.
Our result shows that for $n>1$ the reversibility does not 
generally hold.}

\vskip 0.3cm
{\bf (iii) $\Rightarrow$ (iv).} 
	For a stationary Markov process $x(t)$, the joint 
probability distribution $P(x(t_1),x(t_2),...,x(t_m)$ is uniquely 
determined by the stationary distribution $P_{eq}(x)$ and the 
transition probability $P(y,t|x)$.  By the standard method in
probability, the time reversibility 
implies \cite{QQG}
\[ \int_{\bfR^n}\int_{\bfR^n} \phi(x)P(x,t|y)P_{eq}(y)\psi(y)dxdy
\]
\begin{equation}
    = \int_{\bfR^n}\int_{\bfR^n} \psi(y)P(y,t|x)P_{eq}(x)\phi(x)dxdy
\label{dbr}
\end{equation} 
being valid for arbitrary $\phi, \psi \in D[\bfR^n]$ where $D$ is 
the space of smooth functions with compact supports. 
Note that $P(y,t|x)$ is the fundamental solution to (\ref{FPE}),
\[ \frac{d}{dt}\left(\int_{\bfR^n} P(y,t|x)f(x)dx \right)_{t=0}
      = {\cal L}[f(y)]            \]
where the differential operator 
\begin{equation}
        {\cal L}[f(x)] = \nabla\cdot(\frac{1}{2}A\nabla f + Bxf).
\label{sldo}
\end{equation}
Differntiating both sides of (\ref{dbr}) with respect to $t$ at 
$t=0$, we have 
\begin{equation}
      \int_{\bfR^n} \phi(x){\cal L}[P_{eq}(x)\psi(x)]dx
    = \int_{\bfR^n} \psi(y){\cal L}[P_{eq}(y)\phi(y)]dy.
\end{equation} 
let $f(x)=\phi(x)P_{eq}(x)$ and $g(x)=\psi(x)P_{eq}(x)$,
then $f$ and $g$ are two arbitrary functions in $S$.  Since 
$P_{eq}(x) > 0$,
\begin{equation}
      \int_{\bfR^n} P_{eq}^{-1}(x)f(x){\cal L}[g(x)]dx
    = \int_{\bfR^n} P_{eq}^{-1}(y)g(y){\cal L}[f(y)]dy.
\end{equation} 
Therefore, the operator ${\cal L}$ is symmetric with respect to 
the reciprocal of its stationary distribution $P_{eq}(x)$:
$w(x)=P_{eq}(x)$.  This result is known to physicists.

\vskip 0.3cm
{\bf (iv) $\Rightarrow$ (v).} 
The differential operator ${\cal L}$ in (\ref{sldo}) can also be rewritten as
\[    {\cal L}[f] = \frac{1}{2}\nabla\cdot(A\nabla f) + (\nabla f)Bx + Tr[B]f  \]
where $Tr[B]$ is the trace of the matrix $B$. The statement ($iv$)
\[      \int e^U g{\cal L}[f]dx = \int e^U f{\cal L}[g] dx,    \]
in which the positive $w(x)=e^{-U}$, $f$ and $g$ $\in S$ are arbitrary 
functions, leads to
\[  \int e^U g \left[\frac{1}{2}\nabla\cdot(A\nabla f)
					+(\nabla f)Bx\right] dx
       = \int e^U f \left[\frac{1}{2}\nabla\cdot(A\nabla g)
				+(\nabla g)Bx\right] dx.  \]
Through integration by parts, the first term on the left-hand--side 
(and similarly for the right-hand-side) 
\[   \int e^U g \nabla\cdot(A\nabla f) dx 
           = -\int e^U (\nabla g)A(\nabla f) dx   
		- \int e^U g(\nabla U)A(\nabla f)dx,    \]
and we have
\[ \int e^Ug\left[\frac{1}{2}(\nabla U)A(\nabla f)-(\nabla f)Bx\right]dx
  =\int e^Uf\left[\frac{1}{2}(\nabla U)A(\nabla g)-(\nabla g)Bx\right]dx. \] 
By a simple rearrangement, we have
\[   \int e^U (g\nabla f-f\nabla g)(\frac{1}{2}A\nabla U-Bx) dx=0.  \]
Since $f$ and $g$ are arbitrary, we have  $\frac{1}{2}A\nabla U-Bx =0$ in which
$U=-\log w$.  Therefore
\[     \nabla\log w(x)+2A^{-1}Bx =0,      \]
that is $epr=0$.

\vskip 0.3cm
{\bf (v) $\Rightarrow$ (i).} 
In (\ref{epr}) $A$ is positive definite. Hence the integrand is always
positive, therefore $epr=0$ implies 
$2A^{-1}Bx=-\nabla\log P_{eq}(x)$.  Therefore $A^{-1}B$ is a symmetric 
matrix. Furthermore, $P_{eq}$ is a normal density and hence, 
$P_{eq} \in L^1$ $\Rightarrow$ $A^{-1}B$ is positive definite.
\qquad\end{proof}

\begin{corollary}
A reversible stationary Gaussian solution to (\ref{ndlsde}) has its
covariance matrix $\Xi=\frac{1}{2}B^{-1}A$ where $B$ has to have all 
its eigenvalues being real and positive.  It has a symmetric 
two-time covariant matrix $R(t,t')=e^{-B|t-t'|}\Xi$, 
and its entropy production is zero. 
\label{fdr}
\end{corollary}

\section{Linear Theory for Irreversible Stationary
Processes} 

	In the previous section, we have shown that a symmetric $A^{-1}B$ in 
(\ref{FPE}) is a necessary and sufficient condition for reversibility of 
the stationary solution of (\ref{ndlsde}). The result also implies 
that all the eigenvalues of matrix $B$ are necessarily positive.  In this 
section, we consider the situation when such symmetric condition 
is absent.  One interesting class of problems is when matrix $B$ has
complex eigenvalues.

\begin{definition}
A linear stochastic system characterized by (\ref{ndlsde}) 
and satisfies the statments in Theorem \ref{basic} is called 
reversible.  A system which is not reversible is called irreversible.
\end{definition}

\begin{lemma}
Fokker-Planck equation (\ref{FPE}) with positive definite $A$ but
unrestricted $B$ has an unique fundamental solution.  The solution is 
Gaussian with 
\begin{equation}
           E[x(t)] = e^{-Bt} x^0, \hspace{0.2cm}
    E[\Delta x(t)\Delta x(t)^T] = \int_0^t e^{-Bs}Ae^{-B^Ts}ds,
          \hspace{0.2cm} (\forall t>0),
\label{nonsta}
\end{equation}
where $\Delta x(t)=x(t)-E[x(t)]$. 
\label{vklema} 
\end{lemma}
\begin{proof}
The proof relies on a direct verification of the
Gaussian function with (\ref{nonsta}) as a solution
to equation (\ref{FPE}).  This has been done
many times by physicists \cite{Ke,VanK}. Hence we will not
repeat the lengthy computation.  With the verification of the 
solution, and uniqueness of the fundamental solution to 
(\ref{FPE}), the lemma is proven. \qquad\end{proof}

\begin{theorem}
The necessary and sufficient condition for nonsweeping is
that matrix $B$ has all its eigenvalues with positive
real parts. 
\label{sweeping}
\end{theorem} 

\begin{proof}
Necessity: If the solution to (\ref{FPE}) is 
nonsweeping, then by the Lemma \ref{vklema} and the sweeping 
theorem from Section 1, its fundamental solution has a 
stationary limit $\in L^1$.  Since $x(t)$ has Gaussian distribution
for $t$, the limit distribution is also Gaussiani with finite
variance.  Therefore, 
\begin{equation}
   \lim_{t\rightarrow \infty} 
        E[\Delta x(t)\Delta x(t)^T] < \infty. 
\label{nsg}
\end{equation}
By the general formula for $e^{-Bs}$ \cite{CL} and using 
$A=\Gamma\Gamma^T$, the convergence of (\ref{nsg}) implies that 
all the eigenvalues of $B$ in (\ref{nonsta}) must have positive 
real parts. 

Sufficiency:  If all the eigenvalues have positive 
real parts, the nonstationary Gaussian solution in (\ref{nonsta})
has a unique Gaussian density as its limit when 
$t \rightarrow \infty$.  With the help of the fundamental solution
for equation (\ref{FPE}) and the Markovian property, a stationary 
Gaussian process related to the quantities in (\ref{nonsta}) 
can be constructed.  Hence the solution of (\ref{FPE}) are 
nonsweeping. 
\qquad \end{proof} 

\begin{corollary}
The necessary and sufficient condition for a stationary 
Gaussian process to be irreversible is the matrix $B$ 
in (\ref{ndlsde}) has all eigenvalues
with positive real parts and $A^{-1}B$ is nonsymmetric. 
\label{irr}
\end{corollary}

\section{A Strong Form of Fluctuation-Dissipation Relation}

	We now limit our discussion to nonsweeping situation
with and without reversibility.  We first give
\vskip 0.3cm \noindent
\begin{definition}
The matrix relation among $B$, $\Gamma$ in (\ref{ndlsde}), and the 
stationary covariance $\Xi$ of $x(t)$ 
\[    \Gamma\Gamma^T = B\Xi+\Xi B^T     \]
is called standard fluctuation-dissipation relation.  Actually,
\[    \Xi = \lim_{t\rightarrow\infty} \int_0^t
                e^{-Bs}Ae^{-B^Ts} ds.    \]
\end{definition}

	Combining the above Theorem \ref{basic} and Corollary
\ref{irr}, we immediately have the following corollary.
\begin{corollary}
The standard fluctuation-dissipation relation is a necessary 
but not sufficient condition for the reversibility of system 
(\ref{ndlsde}).  For a reversible system, it can be simplified 
into a stronger form
\begin{equation}
   		A=2B\Xi.
\label{sffdr}              
\end{equation} 
\end{corollary} 
\begin{proof}
By Theorem \ref{basic}, (\ref{ndlsde}) with reversibility has
symmetric $\Gamma^{-T}\Gamma^{-1}B$; and by Corollary \ref{fdr}
its stationary process has covariance
$\Xi = \frac{1}{2}B^{-1}\Gamma\Gamma^T$.  Therefore,
$B\Xi$ = $\frac{1}{2}\Gamma\Gamma^T$ is symmetric, and 
\begin{equation}
      \Gamma\Gamma^T = 2B\Xi = B\Xi + \Xi B^T.    
\label{wffdr}
\end{equation} 
Hence, the standard fluctuation-dissiplatin relation follows,
and furthermore (\ref{wffdr}) can be simplified into 
$A = 2B\Xi$. 

	On the other hand, the standard fluctuation-dissipation
relation is satisfied by any stationary Gaussian processes
with or without the symmetric $A^{-1}B$: From (\ref{nonsta})
we have 
\[     B\int_0^t e^{-Bs}Ae^{-B^Ts} ds + 
 \left(\int_0^t e^{-Bs}Ae^{-B^Ts} ds\right)B^T = 
 -\left[ e^{-Bt}Ae^{-B^Tt} \right]_0^t.           \]
Let $t \rightarrow \infty$, the upper limit vanishes and we have
\[ B\Xi + \Xi B^T = A.   \] 
Hence by Corollary \ref{irr} it is not a sufficient condition for  
reversibility.  \qquad \end{proof}

	(\ref{sffdr}) has a close resemblance to
Einstein's original fluctuation-dissipation relation,
in which $A$ is the covariance of the fluctuating white noise, 
$B$ is the dissipative linear relaxation rates, and $\Xi$ is 
the equilibrium covariance ($kT$). 

\section{The Onsager's Hypothesis and Green-Kubo Formula} 

	The following statement is known as Onsager's hypothesis
or Green-Kubo formula
\cite{On}.  

\begin{theorem}
If a system in (\ref{ndlsde}) is reversible, the conditional 
expectation $E[x(t)|x_0]$ and the two-time covariance of its 
stationary solution have identical time dependence in the
following sense:
\begin{equation}
    E[x(t)|x_0] = e^{-Bt}x_0, \hspace{1cm}
    R(t,0) = e^{-Bt}\Xi,
\label{osh}
\end{equation}  
where $\Xi=E[xx^T]$ is the covariance of the stationary solution.
\end{theorem} 

\begin{proof}
The proof of this result is contained in the proof of Theorem 
\ref{basic}, $(i) \Rightarrow (ii)$.
\qquad\end{proof} 

{\bf Remark:} This is Onsager's original statement \cite{On}. 
However, the statement is not limited to reversible systems,  
it is also applicable to nonsweeping irreversible systems:
\[  R(t,0) = E_{x_0}\left[E[x|x_0]\ x_0^T\right] 
           = E_{x_0}\left[e^{-Bt}x_0 x_0^T\right]  
           = e^{-Bt} \Xi. \]
Indeed, this result which relies solely on the first equality in (\ref{osh}) 
is in fact a consequence of the linearity of equation (\ref{ndlsde}).

\end{article}

\begin{thebibliography}{10} 
\bibitem{CL}
{\sc E.~A. Coddington and N. Levinson}, {\em Theory of Ordinary 
Differential Equations}, McGraw-Hill, New York, 1955.

\bibitem{RRF} 
{\sc R.~R. Fox}, {\em Gaussian Stochastic Processes in Physics},
  Phys. Rep., 48 (1978), pp.~180--283.   

\bibitem{GQW} 
{\sc M.~Z. Guo, M. Qian, and Z.~D. Wang}, {\em The Entropy Production 
  and Circulation of Diffusion Processes on Manifold}, Chin. Sci. 
  Bull., 42 (1998), pp.~982--985.

\bibitem{Hill}
{\sc T.L. Hill}, {\em Free Energy Transduction and Biochemical Cycle
Kinetics}, Springer-Verlag, New York, 1995.

\bibitem{JAP}
{\sc F. J\"{u}licher, A. Ajdari, and J. Prost}, {\em Modeling
Molecular Motors}, Rev. Mod. Phys., 69 (1997), pp~1269--1281.

\bibitem{LM}
{\sc A. Lasota and M.~C. Mackey}, {\em Chaos, Fractals, and Noise: Stochastic
  Aspects of Dynamics}, Springer-Verlag, New York, 1994.  

\bibitem{LS}
{\sc J.L. Lebowitz and H. Spohn},  {\em A Gallavotti-Cohen-Type
Symmetry in the Large Deviation Functional for Stochastic Dynamics},
J. Stat. Phys., 95 (2000), pp~333--365. 

\bibitem{Ke}
{\sc J. Keizer}, {\em Statistical Thermodynamics of Nonequilibrium
  Processes}, Springer-Verlag, New York, 1987. 
 
\bibitem{Ok}
{\sc B.~K. \O ksendal}, {\em Stochastic Differential Equations:
   An Introduction with Applications}, third Ed., Springer-Verlag,
   New York, 1997. 

\bibitem{On}
{\sc L. Onsager}, {\em Reciprocal Relations in Irreversible
Processes. I}, Phys. Rev., 37 (1931), pp.~405--426. 

\bibitem{Qian}
{\sc H. Qian}, {\em Vector Field Formalism and Analysis for
  a Class of Thermal Ratchets}, Phys. Rev. Lett., 81 (1998),
  pp.~3063--3066. 

\bibitem{Qian2}
{\sc H. Qian}, {\em Single-Particle Tracking: Brownian Dynamics 
of Viscoelastic Materials}, Biophys. J., 79 (2000), pp.~137--143.

\bibitem{Qian3}
{\sc H. Qian}, {\em A Mathematical Analysis for the
Brownian Dynamics of DNA Tether}, J. Math. Biol., 41 (2000), 
pp.~331--340.

\bibitem{Qian4}
{\sc H. Qian}, {\em The Mathematical Theory of Molecular Motor 
Movement and Chemomechanical Energy Transduction},
J. Math. Chem., in the press.  

\bibitem{Qian5}
{\sc H. Qian}, {\em Equations for Stochastic Macromolecular Mechanics 
of Single Proteins: Equilibrium Fluctuations, Transient Kinetics and 
Nonequilibrium Steady-State}, J. Chem. Phys., submitted.  

\bibitem{Qian6}
{\sc H. Qian}, {\em Nonequilibrium Steady-State Circulations
and Heat Dissipation Functional}, Phys. Rev. E., in the press.

\bibitem{Qian7}
{\sc H. Qian}, {\em Relative Entropy: Free Energy Associated with 
Equilibrium Fluctuations and Nonequilibrium Deviations}, 
Phys. Rev. E., in the press.

\bibitem{QW}
{\sc M. Qian and Z.~D. Wang}, {\em The Entropy Production of
  Diffusion Processes on Manifolds and Its Circulation 
  Decomposition}, Commum. Math. Phys., 206 (1999), pp.~429--445.  

\bibitem{QQ}
{\sc M.~P. Qian and M. Qian}, {\em The Entropy Production and 
  Irreversibility of Markov Processes}, Chin. Sci. Bull., 30 
  (1985), pp.~445--447.

\bibitem{QQG}
{\sc M.~P. Qian, M. Qian, and G.~L. Gong}, {\em The Reversibility and
  the Entropy Production of Markov Processes}, Contemp. Math., 118 (1991),
  pp.~255--261.

\bibitem{VanK}
{\sc N.~G. Van Kampen}, {\em Stochastic Processes in Physics and
   Chemistry},  Revised and enlarged Ed., North-Holland, Amsterdam, 1997. 

\bibitem{WU}
{\sc M.~C. Wang and G.~E. Uhlenbeck}, {\em On the Theory of the 
  {B}rownian Motion II}, Rev. Mod. Phys., 17 (1945), 
  pp.~323--342.

\bibitem{Wax}
{\sc N. Wax}, {\em Selected Papers on Noise and Stochastic Processes},
   Dover, New York, 1954.  

\bibitem{Weiss}
{\sc G. Weiss}, {\em Time-Reversibility of Linear Stochastic Processes},
  J. Appl. Prob., 12 (1975), pp.~831--836. 

\end{thebibliography}
\end{document}